# Light switching based on half space invisible states

A. R. BEKIROV, A. F. USPENSKIY, V. A. SITNYANSKY, B. S. LUK'YANCHUK AND A. A. FEDYANIN

## Abstract

We investigate a method for controlling light scattering based on the excitation of non-radiating states in a half-space through a tailored choice of incident radiation. For a fixed particle geometry, we demonstrate that small variations in the refractive index can lead to a significant redistribution of scattered light between two half-spaces while keeping the incident illumination unchanged. This effect is particularly relevant for dynamic beam shaping and optical signal routing at the microscale. Our study focuses on spherical semiconductor particles of varying radii illuminated in the visible range, with refractive index modulation achieved via charge carrier injection. Using AlGaAs and InP as model materials, we analyze the feasibility of achieving efficient directional control of scattering. These results provide insight into all-optical manipulation of light using tunable semiconductor structures.

## 1. Introduction

The rapid development of photonic technologies in recent years has been fueled by the growing demand for faster, more efficient, and miniaturized optical devices. Among these advancements, all-optical switching and light-induced refractive index modulation have emerged as promising phenomena, offering unprecedented opportunities for ultrafast data processing, optical communication, and photonic computing systems.

All-optical switching eliminates the need for electrical control, relying entirely on the interaction between light and matter to achieve state transitions. This approach enables operational speeds on the order of femtoseconds and paves the way for compact, energy-efficient photonic circuits. Additionally, the ability to dynamically modulate the refractive index under intense light exposure serves as the foundation for nonlinear optics, facilitating applications such as self-focusing, optical limiting, and waveguide tuning.

At the core of these processes lie nonlinear optical effects, including the Kerr effect, multiphoton absorption, and optically induced phase transitions. A comprehensive understanding of this modulation requires consideration of carrier generation dynamics under radiation exposure. For example, the ABC model has been employed in various works [1, 2]. A more accurate description involves solving the Maxwell-Bloch equations [3], where the optical modulation of material properties is inherently accounted for through the dynamic response of the medium.

**In this work, we adopt a simplified approach, assuming fixed carrier populations in the ground and excited states.**

These effects are critically influenced by the intrinsic properties of materials, such as their nonlinear susceptibility, damage threshold, and response time. While traditional materials like silicon and gallium arsenide have been instrumental in advancing these studies, emerging materials such as photonic crystals, organic polymers, and metamaterials are broadening the scope of light-matter interactions. In our study, we use **AlGaAs** and **InP** as the materials of interest.

Typically, studies assume a planar wavefront for the incident light. However, in this work, we deviate from this assumption to investigate more complex scenarios. This paper aims to analyze the mechanisms underlying optical switching based on the excitation of half-space invisible states [4]. Our approach focuses on optimizing both the geometry of the material and the type of excitation radiation.

## 2. Effect of Carrier Concentration on the Dispersion of Refractive Index and Absorption in AlGaAs and InP Materials

The change in the complex refractive index that occurs due to the injection of free charge carriers into the volume of the semiconductor can be described within the quantum model by three processes [5]: absorption by free carriers, bandfilling shrinkage, and filling of bands (Burstein-Moss effect).

Free carrier absorption, which involves intraband transitions within the conduction band, contributes to the refractive index change according to the Drude model as follows:

$$\Delta n_{FCA} = -\frac{e^2\lambda^2}{8\pi^2 c^2 \varepsilon_0 n}\left(\frac{N}{m_e}+\frac{P}{m_h}\right).$$

Bandgap shrinkage occurs due to electron interactions at the conduction band edge and the Pauli exclusion principle, which lead to a lowering of the conduction band minimum (similarly for holes).

In contrast, band filling increases the effective bandgap, as the lower energy states in the conduction band become occupied. As a result, electrons require higher energies for transitions, exceeding the intrinsic bandgap energy $E_g$.

These two processes contribute to a change in the optical absorption coefficient, $\Delta\alpha_{BF,BS}(E)$, which is used to calculate the modification of the imaginary part of the refractive index:

$$\operatorname{Im}\Delta n_{BF,BS}(E) = \lambda\, \Delta\alpha_{BF,BS}(E)/4\pi,$$

while the real part is derived using the Kramers–Kronig relations:

$$Re\Delta n_{BF,BS}(E) = \frac{ch}{\pi e^2}\int_0^\infty \frac{\Delta\alpha_{BF,BS}(E')}{(E'^2 - E^2)} dE'.$$

Near the bandgap energy, the absorption coefficient for a direct-bandgap semiconductor can be approximated by the following relation:

$$\alpha(E) = C/E\,\sqrt{E-E_g}\cdot(E>E_g).$$

Its variation due to band filling is calculated using the Fermi–Dirac distribution:

$$\Delta\alpha_{BF}(E) = \alpha(E)\big(f(E_v) - f(E_c) - 1\big),$$

where $f(E) = (1 + exp[(E_F - E)/k_B T])^{(-1)}$ is the Fermi distribution, and $E_v$ and $E_c$ are the energies of the valence band and conduction band edges, respectively. Bandgap shrinkage reduces the effective bandgap width by $\Delta E_g$ (see [5, 6]).

Taking both processes into account, the total change in the absorption coefficient is given by [7]:

$$\Delta\alpha_{BF,BS}(E) = \alpha(E - \Delta E_g)(f(E_v) - f(E_c)).$$

The contributions from electrons and holes should be calculated separately [5]. The total change in the refractive index is obtained as the sum of the individual contributions:

$$n(N,P) = n_0 + \Delta n_{BF,BS} + \Delta n_{FCA}. \quad (1)$$

Figure 1 shows the dispersion dependence of the refractive index and absorption coefficient for different carrier concentrations: N=$10^{18}$ $cm^{-3}$ and N=$10^{19}$ $cm^{-3}$ for two materials, AlGaAs and InP.

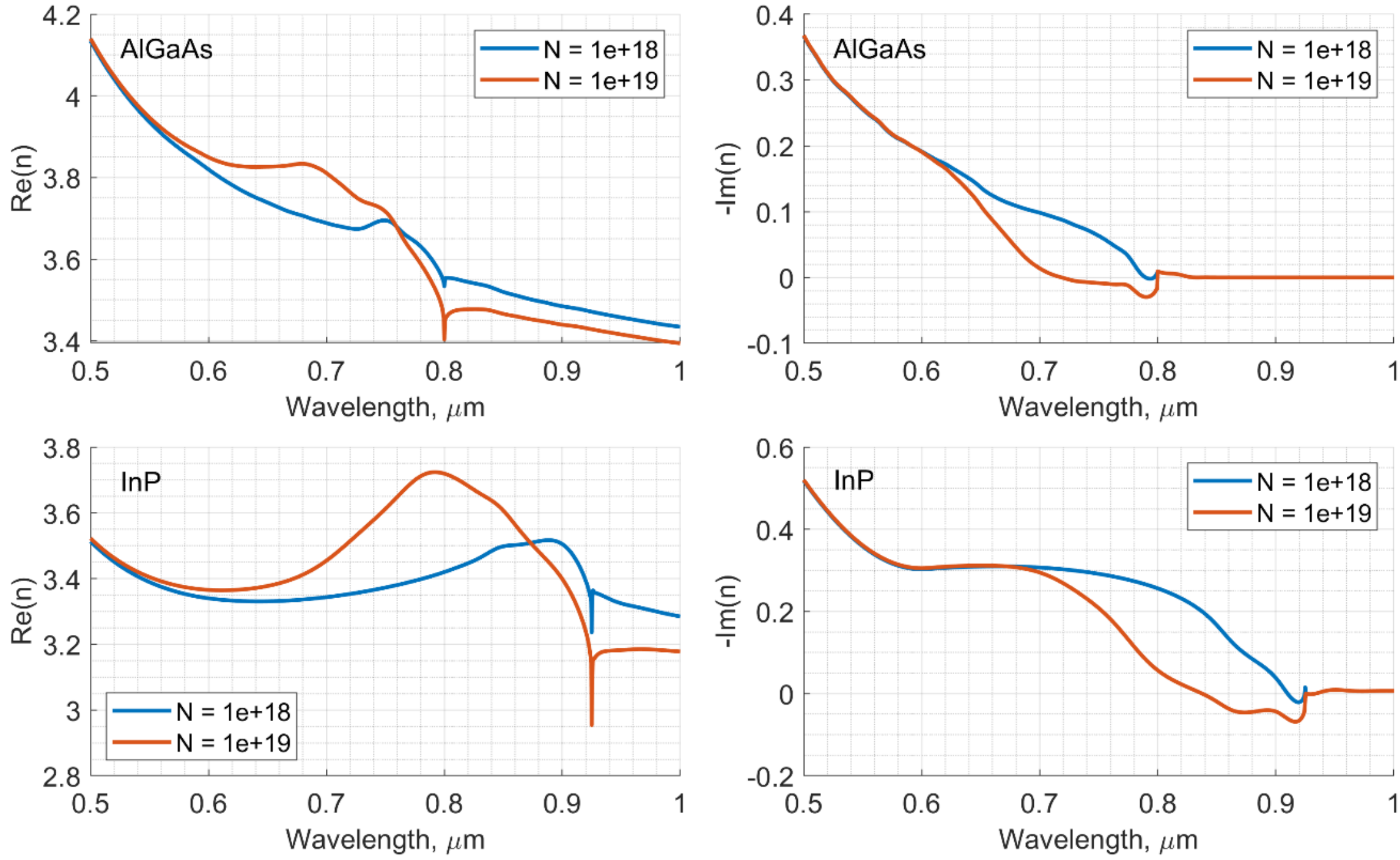

Fig. 1 Dispersion dependence of the refractive index for different carrier concentrations: N=$10^{18}$ $cm^{-3}$ and N=$10^{19}$ $cm^{-3}$.

## 3. Scattering of Electromagnetic Waves by a Particle: T-Matrix Formalism and Control of Radiation Patterns

The scattering of electromagnetic waves by a particle can be effectively described using the T-matrix formalism [8], which establishes a relationship between the incident field and the field scattered by the particle. In this work, we focus on homogeneous particles with a constant refractive index at a specific wavelength of light.

Both the incident field ($\mathbf{E}^{inc}$) and the scattered field ($\mathbf{E}^{sca}$) can be expressed in terms of expansion coefficients ($g_{lm}$, $f_{lm}$) and ($b_{lm}$, $a_{lm}$), respectively. These coefficients are derived from the representation of the fields using basis functions, commonly chosen as the vector spherical harmonics $\mathbf{M}_{lm}$ and $\mathbf{N}_{lm}$ [9]:

$$\mathbf{E}^{inc} = \sum_{l=1}^{l_{\max}} \sum_{m=-l}^{l} f_{lm}\mathbf{N}_{lm}^{(1)} + g_{lm}\mathbf{M}_{lm}^{(1)}, \tag{2}$$

$$\mathbf{E}^{sca} = \sum_{l=1}^{l_{\max}} \sum_{m=-l}^{l} a_{lm}\mathbf{N}_{lm}^{(3)} + b_{lm}\mathbf{M}_{lm}^{(3)}. \tag{3}$$

where $l_{\max}$ is the maximum number of excited modes, which is determined by the convergence condition of series (2)–(3). The functions **N** and **M** are given by:

$$\mathbf{M}_{lm}^{(1,3)} = z_l(\rho)e^{im\varphi}(i\pi_{lm}\mathbf{e}_\theta - \tau_{lm}\mathbf{e}_\varphi),$$

$$\mathbf{N}_{lm}^{(1,3)} = \rho^{-1}z_l(\rho)l(l+1)P_l^m(\cos\theta)e^{im\varphi}\mathbf{e}_r + \rho^{-1}[\rho z_l(\rho)]' e^{im\varphi}(\tau_{lm}\mathbf{e}_\theta + i\pi_{lm}\mathbf{e}_\varphi),$$

where the superscript (1) indicates that $z=j_l$, and (3) $z=h_l$, $j_l$ and $h_l$ are the spherical Bessel function and the Hankel function of the first kind, respectively, $\rho=kr$, $k$ is the wavenumber in the medium, $\pi_{mn}(\cos\theta) = mP_l^m(\cos\theta)/\sin\theta$, $\tau_{mn}(\cos\theta) = dP_l^m(\cos\theta)/d\theta$ and $P_l^m$ associated Legendre polynomials and $(r, \theta, \varphi)$ are spherical coordinates with the origin at the center of the particle. The T-matrix connects the columns of the incident field coefficients $E^i = (g_{lm}, f_{lm})$ and the scattered field coefficients $E^s = (b_{lm}, a_{lm})$, where the indices $l$ and $m$ span all possible values. If the particle is axisymmetric relative to the z-axis, the coefficients with different $m$-indices can be calculated independently:

$$E_m^s = T_m E_m^i. \tag{4}$$

In this relation, in the vector columns $E^i_m$, $E^s_m$, the index $m$ is fixed, while the index $l$ spans all possible values, i.e., $l=\max(|m|,1), |m|+1, |m|+2,\ldots, l_{\max}$. Relation (4) holds for each azimuthal mode $m$ separately.

In the case where the distance from the particle is much greater than the wavelength, i.e., $r \gg \lambda$, the scattered field can be expressed as:

$$\mathbf{E}^{sca} = \mathbf{F}(\theta,\varphi)\exp(ikr)/(ikr) \tag{5}$$

Where **F** is the scattering amplitude. The scattering amplitude **F** can be expressed in terms of the expansion coefficients introduced in Eq. (3). This expression is conveniently written using two auxiliary functions, $S_1$ and $S_2$:

$$\mathbf{F}(\theta,\varphi) = S_1\mathbf{e}_\theta + S_2\mathbf{e}_\varphi, \tag{6}$$

The functions $S_1$ and $S_2$ are defined as follows:

$$S_1 = -\sum\nolimits_{l,m} \exp(im\varphi)(-i)^{l+1}[a_{lm}\tau_{lm} + b_{lm}\pi_{lm}],$$
$$S_2 = -\sum\nolimits_{l,m} \exp(im\varphi)(-i)^{l}[a_{lm}\pi_{lm} + b_{lm}\tau_{lm}]. \tag{7}$$

We aim to achieve a sharp change in the radiation pattern **F** by varying the refractive index, as determined by the dependency (1). Our approach for controlling the radiation pattern involves exciting non-radiating in half space states in the particle through incident radiation. A particle is considered to be in a non-radiating state if the scattering coefficient vector $E^s_m$ is a linear combination of the columns of a certain matrix $Q_m$. In this case, the scattering amplitude satisfies the following relationship:

$$\mathbf{F}(\theta,\varphi) \equiv 0, \text{at } \theta \le \pi/2. \text{ If } E_m^s = Q_m E_m^Q, \tag{8}$$

where $E^Q{}_m$ is an arbitrary vector of coefficients describing the linear combination. The matrix of half-space non-radiating states, $Q_m$, was first computed in [10], and this concept was further developed in [4].

Each column of the matrix $Q_m$ represents a decomposition of the type (3) corresponding to the mode $\mathbf{N}^{(3)}_{l_0 m_0}$ or $\mathbf{M}^{(3)}_{l_0 m_0}$, in which the spatially propagating harmonics are omitted. In this case, the coefficients $a_{lm}$, $b_{lm}$ contain only modes with the same azimuthal number $m=m_0$. The orbital modes $l$ include the mode with $l=l_0$, as well as modes with opposite parity, which decay as $|l-l_0|$ increases.

To ensure that condition (8) is satisfied with high accuracy, if the column $E^s{}_m$ contains $l_{\max}$ modes, then the maximum mode number $l_0=l_{\max}{}^Q$ must be less than $l_{\max}$. In our work, we chose $l_{\max}{}^Q = l_{max}$-8. Therefore, if $l_{\max} = 10$, for $m$=1, the dimensions of the matrix $T_{m=1}$ will be 10×10, the dimensions of $E^s{}_m$ and $E^i{}_m$ will be 10×1, and the dimensions of the matrix $Q_{m=1}$ will be 10×2, while the column $E^Q{}_m$ will be 2×1.

If we consider $E^s{}_m$ as a linear combination of the complex conjugate matrix $(Q_{-m})^*$ with the opposite azimuthal index $m$, then the condition in Eq. (8) will hold in the opposite half-space, i.e., for $\theta \geq \pi/2$. This means that the particle will be non-radiating in the opposite direction.

## 4. Rapid light switching and excitation of invisible states

Our goal is to provide an example of a surrounding field in which, at a refractive index $n_1$, the particle emits radiation in one half-space, and at $n_2$, it emits in the opposite half-space. According to this setup, for $n=n_1$, the condition $E^s{}_{1,m}=Q_m E^Q{}_{1,m}$ should be satisfied, and for $n=n_2$, $E^s{}_{2,m}=(Q_{-m})^* E^Q{}_{2,m}$ holds. In this section, we consider the case in which $m$=1 and all other modes are absent. For this reason, we will omit the index $m$ in the notation. Using the relation in Eq. (8), these conditions can be written as:

$$E_1^s = T_1 E^i = Q E_1^Q\,, \tag{9}$$

$$E_2^s = T_2 E^i = (Q)^* E_2^Q\,, \tag{10}$$

where $T_1$ – scattering matrix of the particle at $n = n_1$, $T_2$ at $n = n_2$. Thus, if the incident field $E^i$ is a solution, it simultaneously excites non-radiating states in the particle in opposite directions. Furthermore, analogous conditions can be formulated for any azimuthal mode $m$.

Nontrivial solutions of equations (9)–(10) do not always exist; for example, when $T_1=T_2$. In that case, these equations reduces to

$$Q E_1^Q = (Q)^* E_2^Q\,,$$

which can only be satisfied if $E_1^Q = E_2^Q = 0$, due to the linear dependence between the even and odd rows of the matrix $Q$ [10]. If the odd rows of $QE^Q{}_1$, are given by $x_1$, $x_2$, $x_3$,…, then the even rows must be uniquely determined by them and are equal to $y_2$,$y_4$,$y_6$,…. However, it is evident that for the left-hand side, due to the same linear dependence, these values will be different—say, $y_2'$, $y_4'$, $y_6'$,…—which implies that $Q E_1^Q \neq (Q)^* E_2^Q$.

Another extreme case occurs when $T_1=Q$ and $T_2=Q^*$; in this situation, due to the relation $Q^2=Q$, (see [10]) a trivial solution can be obtained by setting $E^Q{}_1= E^Q{}_2= E^i$, where $E^i$ can be chosen arbitrarily.

The intermediate cases, in which $T_1\neq T_2$, are more complex. The question of the existence of a solution and its determination is nontrivial and can be approached in various ways. In every case, one must work with the two columns $E^Q{}_1$ and $E^Q{}_2$, which should be chosen in the most advantageous way from the standpoint of ensuring a solution exists.

We tried various approaches to finding solutions in the general case. The most effective method turned out to be one that takes advantage of the linear dependence between the even and odd rows of the matrices $Q$ and $Q^*$. Because of this dependency, in equations (9)–(10) it is sufficient to equate only the even (or odd) rows rather than all rows, while simultaneously choosing the arbitrary columns $E^Q{}_1$ and $E^Q{}_2$ such that they are as close as possible to the corresponding vectors $T_1E^i$ and $T_2E^i$ in terms of the root mean square deviation. We will explain our idea step by step.

Let the scattered field vector $E^s{}_1$ be given. We choose the product $QE^Q{}_1$ such that the norm of the difference between these vectors is minimized. This problem is solved by using the pseudoinverse matrix, i.e.,

$$E_1^Q = Q^\dagger E_1^s ,$$

where $Q^\dagger$ is a pseudoinverse of $Q$. This choice ensures that the difference $|E^s{}_1-QE^Q{}_1|$ is minimized. Similarly, one can choose

$$E_2^Q = \left(Q^*\right)^\dagger E_2^s .$$

In this case, equations (9)–(10) can be rewritten as:

$$QQ^\dagger T_1 E^i = T_1 E^i \quad ,$$

$$Q^*\left(Q^*\right)^\dagger T_2 E^i = T_2 E^i \quad .$$

Transferring the left-hand side of the equations to the right-hand side:

$$(QQ^\dagger - I)T_1 E^i = 0\,, \tag{11}$$

$$(Q^*\left(Q^*\right)^\dagger - I)T_2 E^i = 0\,. \tag{12}$$

The final step involves isolating the even and odd rows. Due to the linear dependence, if $T_1E^i$ lies within the linear space of $Q$ and the even (or odd) rows are equal to the even (or odd) rows of $QE^Q{}_1$, then the entire vector $T_1E^i$ is equal to $QE^Q{}_1$, i.e., $T_1E^i$= to $QE^Q{}_1$. Similarly, this holds true for $T_2E^i$ and $Q^*E^Q{}_2$. Thus, if a solution to the system (11)–(12) exists, the number of equations can be halved by keeping only the even rows in (11) and the odd rows in (12). In this case, the system transforms into a square form:

$$\left(QT\right)_{1/2} E^i = 0\,, \tag{13}$$

where the even rows of matrix $(QT)_{1/2}$ are equal to the even rows of matrix $(QQ^\dagger - I)T_1$, and the odd rows of matrix $(QT)_{1/2}$ are equal to the odd rows of matrix $(Q^*\left(Q^*\right)^\dagger - I)T_2$. Next, it is necessary to find the nontrivial eigenvectors of matrix $(QT)_{1/2}$ with zero eigenvalues λ:

$$\left(QT\right)_{1/2} E^i_\lambda = \lambda E^i_\lambda \tag{14}$$

The search for eigenvalues was performed numerically, after which values were selected for which the equality holds:

$\lambda_j^{zero} = \{|\lambda_i| \lhd 10^{-4}, i = 1,2,3...\}$.

If a solution exists, the found eigenvector will be a solution in the sense of the best approximation. For this reason, all found eigenvectors must be verified to satisfy the conditions of equations (9)–(10). **The case of the absence of solutions will be discussed in more detail below.**

Since the goal of the work is light switching, in addition to the condition that the scattering amplitudes are equal to zero in the corresponding half-spaces, we aim for the condition that the maximum values of the amplitude functions are equal, to ensure good contrast in the switching. Furthermore, we seek solutions where the maximum of the incident field is concentrated within the volume of the particle. Otherwise, only a small portion of the incident radiation would interact with the particle. Thus, the desired conditions for a "good" solution are:

1. The scattered field $E^s{}_1$ does not radiate in the lower half-space (z<0), and the scattered field $E^s{}_2$ does not radiate in the upper half-space (z>0).
2. The maximum of the scattering amplitude $|\mathbf{F}_1|$ is equal to the maximum of the amplitude $|\mathbf{F}_2|$.
3. The maximum value of the incident field $\boldsymbol{E}^{inc}$ is concentrated within the volume of the particle.

Condition 3 is satisfied due to the limited number of modes in the incident field expansion (2). According to the localization principle [11], the term of order $l$ corresponds to a ray passing at a distance of $(l+1/2)\lambda/2\pi$ from the origin. When $l+1/2=2\pi R/\lambda=q$, this distance exactly equals the radius of the sphere R, and such terms describe waves that effectively interact with the particle. Terms with $l+1/2<q$ correspond to rays falling on the sphere and describe diffraction and scattering processes inside the particle. Thus, we chose the number of modes considered in the decomposition of $\mathbf{E}^{inc}$ in Eq. (2):

$$l_{\max} = \begin{cases} fix(q) - 5, if\ fix(q) - odd \\ fix(q) - 6, if\ fix(q) - even \end{cases},$$

where fix($q$) denotes the nearest integer less than $q$. We choose different offsets in modes 5 and 6 for even and odd fix($q$) to ensure that $l_{\max}$ is always even, which simplifies the calculations.

Conditions (1) and (2) are not necessarily satisfied during the solution process and therefore require additional verification. To evaluate the efficiency of the switching process, it is useful to introduce a quantitative parameter. For this purpose, we define a quantity that characterizes the excitation efficiency of the Q modes:

$$S^Q = \frac{I^{(forw)} - I^{(back)}}{I} \tag{15}$$

Where $I = \iint |\mathbf{F}| d\Omega$, $I^{(forw)} = \iint_{\theta \geq \pi/2} |\mathbf{F}| d\Omega$, $I^{(back)} = \iint_{\theta < \pi/2} |\mathbf{F}| d\Omega$. The parameter $S^Q$ ranges from -1 to 1. When $S^Q$=−1, all scattered radiation is directed into the half-space z<0, indicating that *Q*-type modes are predominantly excited within the particle. Conversely, when $S^Q$=+1, the scattering is directed into the half-space z>0, corresponding to the excitation of $Q^*$-type modes. To quantify the optical switching between these two states, while taking into account the requirement of equal scattering amplitudes, we define the parameter as follows:

$$S = \max_{\lambda_j^{zero}} \left( \frac{1}{2} \frac{S_1^Q max(|\mathbf{F}_1|) - S_2^Q max(|\mathbf{F}_2|)}{max(|\mathbf{F}_1|, |\mathbf{F}_2|)} \right), \quad (16)$$

The quantity S varies from –1 to 1 and characterizes the efficiency of radiation redirection for different refractive indices $n_1$ and $n_2$. When S=−1, the particle emits entirely into the half-space z<0 in both states. When S=0, the radiation is equally distributed in both directions, and no redirection occurs. When S=1, the radiation in the two states is directed into opposite half-spaces with equal maximum amplitudes. Thus, S=1 corresponds to the ideal case of optical switching in the chosen metric. In addition, in Eq. (16), the best solution obtained from Eq. (14) is selected among all $\lambda_j^{zero}$ values.

Fig. 2 shows the optical switching efficiency S as a function of wavelength and particle radius. The refractive index variation is defined by Eq. (1) or by the dependencies presented in Fig. 1.

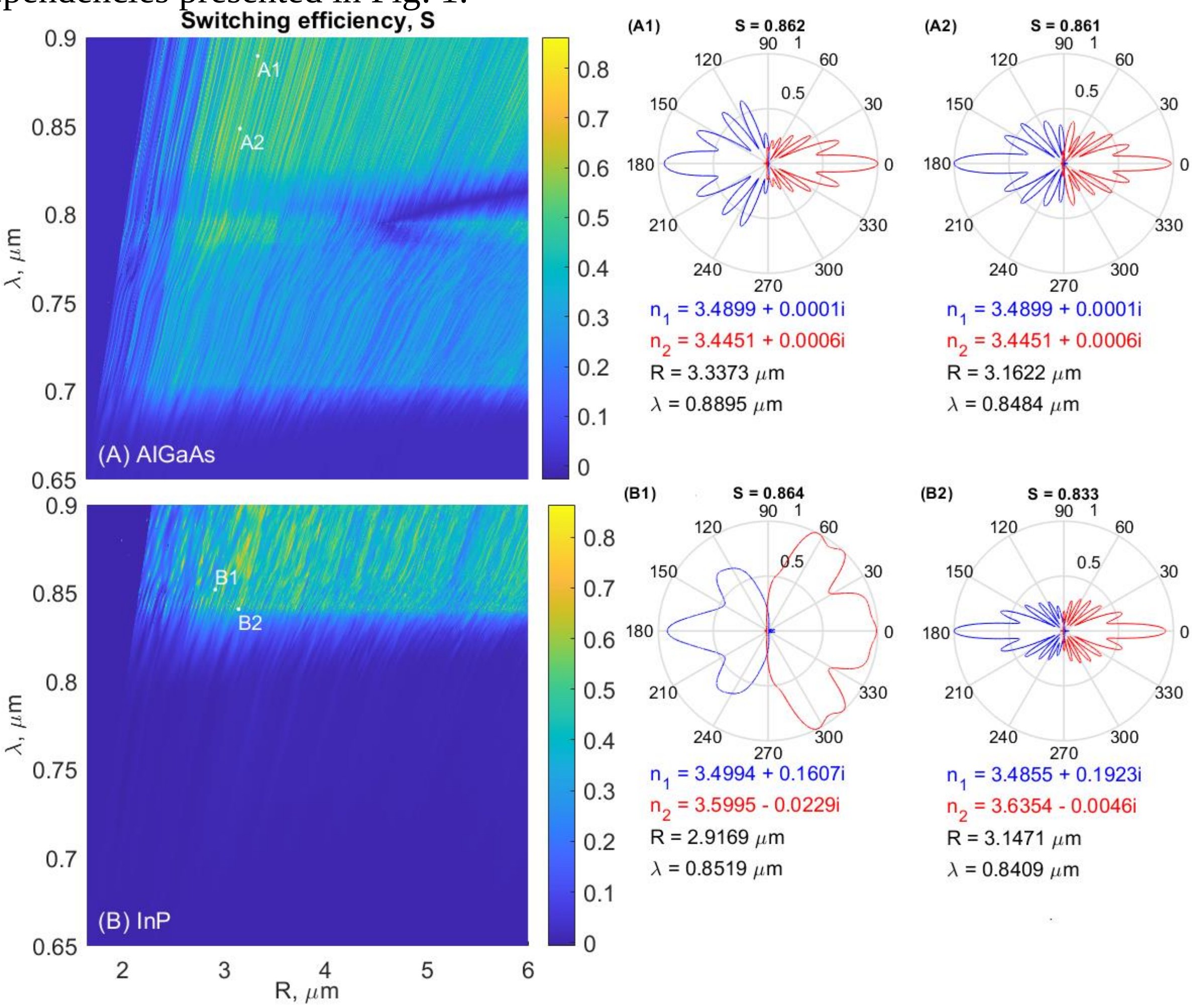

**Fig. 2.** The optical switching efficiency parameter in both the normal and excited states as a function of wavelength and particle size. (A) – The general distribution of the S-parameter for AlGaAs, (B) – for InP. (A1, A2)–scattering amplitude moduli for AlGaAs for two best S-parameter values: 0.862, 0.861, respectively. (B1, B2) – similarly for InP, with S-values: 0.864, 0.833, 0.78584, 0.5. The blue curves correspond to scattering at N=18 $cm^{-3}$, and the red curves correspond to N=19 $cm^{-3}$. The positions of the points are marked on the graphs with white markers and corresponding labels. The highest value of the S-parameter is achieved at points A1 and B1.

It is interesting to note that the threshold for exciting efficient light scattering switching differs between the two materials. For AlGaAs, this threshold occurs around 0.7 μm, while for InP, it starts at approximately 0.85 μm. These values are determined by the dispersion dependence (1) shown in Fig. 1, as noticeable changes in material properties begin after a certain limit. Another characteristic feature is the distinct dip for AlGaAs around 0.8 μm and the values of R between 4.5 and 6 μm. We have observed this behavior consistently across different approaches to solving equations (9)-(10), indicating that this is a common property of the material.

Another notable feature is the clear boundary for small sizes (R<1.8 μm), where no effective light switching occurs. This boundary can shift due to different choices of $l_{max}$. This happens because the number of modes considered in the particle is insufficient to excite at least one non-radiating state, which results in no solutions of Eq. (13). In such cases, we assume S=0.

The best found values for the S-parameter are as follows: for AlGaAs, $S_{max}$=0.862 at λ=0.8895 μm, R=3.3373 μm; for InP, $S_{max}$=0.864 at λ=0.8519 μm, R=2.9169 μm. Fig. 3 shows the incident and scattered fields of the particle at different refractive indices, corresponding to the point of maximum switching efficiency.

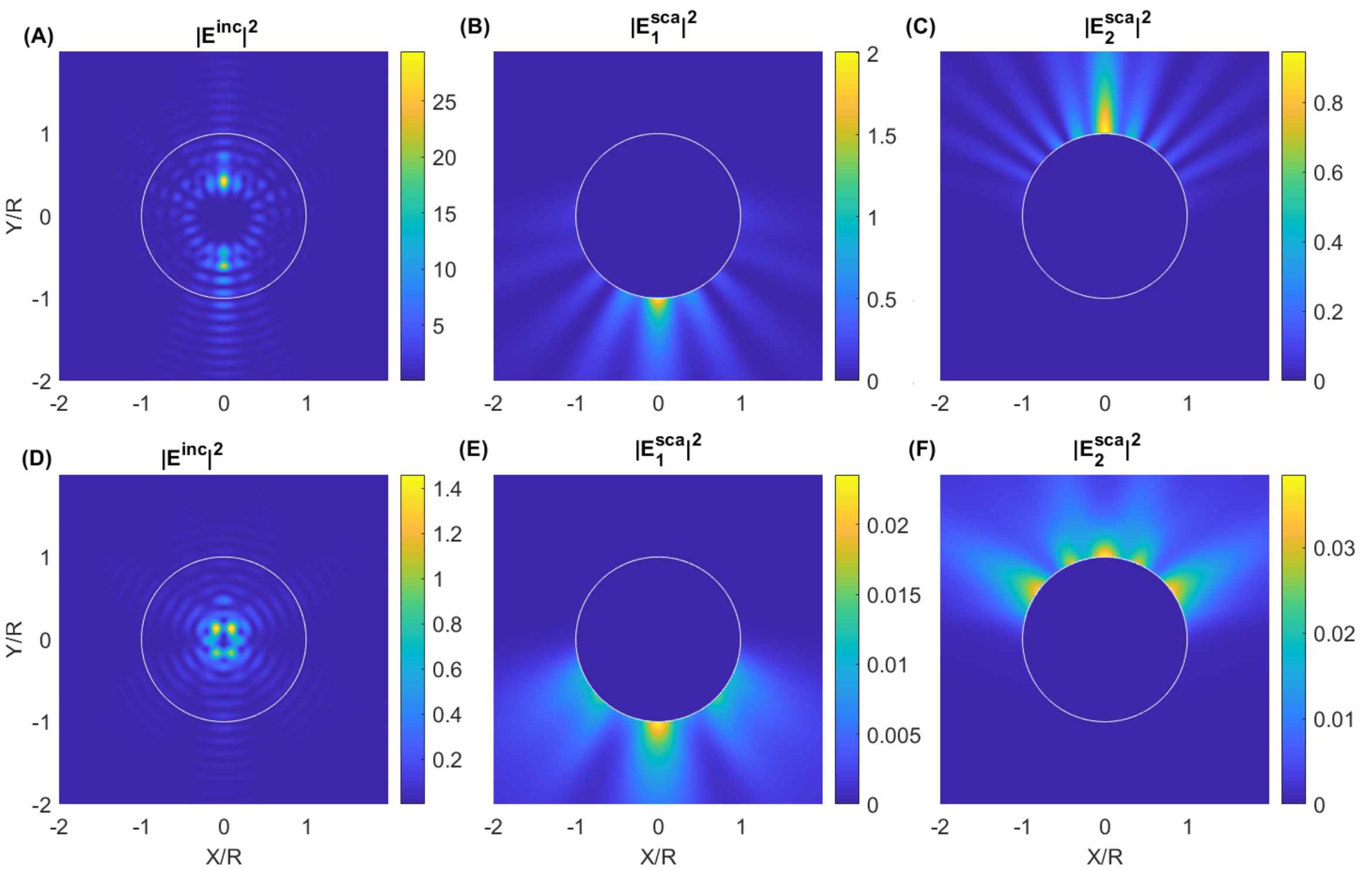

Fig. 3. (A) – Incident radiation on the particle. (B) - Scattered field at N=$10^{18}$ $cm^{-3}$. (C) - Scattered field at N=$10^{19}$ $cm^{-3}$, for AlGaAs. Radiation and particle parameters: λ=0.8895 μm, R=3.3373 μm,

$S=S_{max}=0.862$ (point A1 on Fig. 2A). (D) - (F) Similar quantities for InP. Radiation and particle parameters: λ=0.8519 μm, R=2.9169 μm, $S=S_{max}=0.864$ (point B1 on Fig. 2B).

## 5. Discussion

Fig. 2(A)-(B) exhibits pronounced oscillations, since at each point a new external field configuration is selected. To accurately estimate the actual resonance width, the external field must be fixed. As an example, let us consider the accuracy of the system parameters at the point of maximum switching, labeled $A_1$, for AlGaAs. Fig. 4(A) presents the switching efficiency as a function of the size parameter q=2πR/λ, calculated for a fixed incident field $\mathbf{E}^{inc}$.

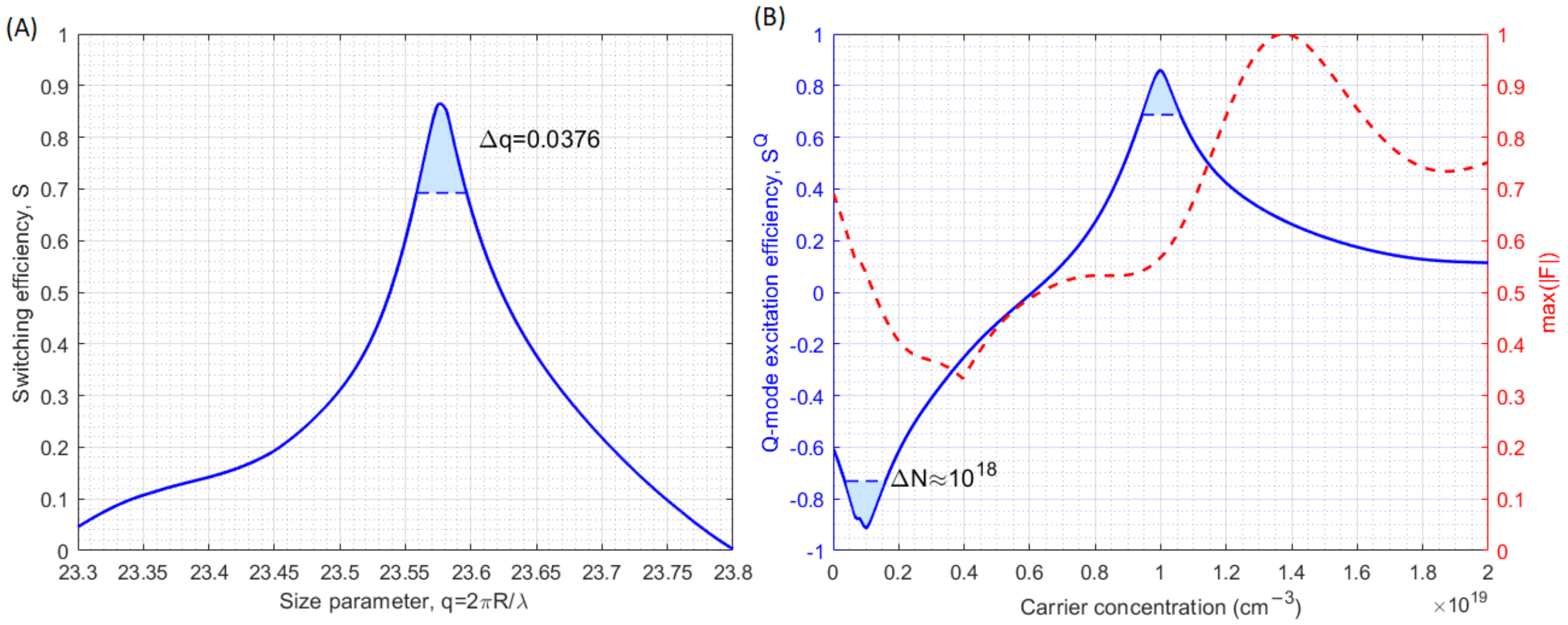

**Fig. 4.** (A) Optical switching efficiency S versus the size parameter q=2πR/λ for a particle in a fixed external field $\mathbf{E}^{inc}$. (B) *Q*-mode excitation parameter $S^Q$ as a function of carrier concentration N. The red dashed curve indicates the normalized scattering amplitude maximum |**F**|. Results are shown for AlGaAs near the resonance point $A_1$ in Fig. 3(A).

The width of the considered resonance, determined as the range where the parameter S deviates by no more than 80% from its maximum value, is estimated to be Δq=2π Δ(R/λ)≈0.04. This corresponds to a wavelength variation of approximately Δλ~1.5 nm for a fixed particle radius R, which is readily achievable with a wide class of laser sources.

Let us now consider the dynamics of the radiation pattern evolution when the carrier density N deviates from its limiting values $N_1$ and $N_2$. The incident field, particle size, and wavelength are assumed to be fixed. The parameter S characterizes the optical switching between two states; however, to analyze the dynamics of the radiation pattern evolution, it is convenient to introduce the parameter $S^Q$. Fig. 4(B) shows the dependence of $S^Q$ on the carrier density N. The ideal switching scenario corresponds to the points where $S^Q$ changes from -1 to 1 under the condition of equal maximum field amplitudes **F**. In this case, the overall switching efficiency parameter S reaches unity. In the considered example, the best value of S=0.85 is obtained for carrier densities $N_1=10^{18}$ $cm^{-3}$ and $N_2=10^{19}$ $cm^{-3}$. As can be seen from the graph, this value remains nearly constant over relatively broad ranges of $N_1$ and $N_2$.

In the present analysis, the refractive index of the particle is assumed to vary uniformly. If the refractive index changes nonuniformly, the above considerations remain valid; however, the scattering matrices $T_1$ and $T_2$ in Eqs. (9)–(10) will have different elements. The existence of such solutions in the nonuniform case requires a separate analysis, but no fundamental limitations arise in this respect.

To implement the proposed concept of light-scattering control, two additional tasks must be addressed: (i) generating the required carrier density N within the particle and (ii) constructing a resonant structure or optical system that supports the necessary spatial modes. Both tasks are nontrivial but can be realized using different approaches.

The carrier density N can be controlled either by electrical injection or by optical excitation. Optical control can be achieved through interband absorption in semiconductors; however, the carrier lifetime and the efficiency of generation and recombination strongly depend on the material properties. For stable control, it is essential to take into account the characteristic carrier relaxation times and heating effects. Various approaches to achieving this have been discussed in the literature [1].

To realize the required spatial field distribution, an optical system comprising at least two lenses placed on opposite sides of the particle is required, since the incident field $\mathbf{E}^{inc}$ is generally non-unidirectional. These lenses can be positioned in the far field, as the desired field contains no evanescent spatial harmonics due to its representation as a series of spherical functions of the first kind. The simplest technique for generating the required fields is to project the Fourier spectrum of the desired radiation onto the lens aperture, so that the lens forms its Fourier transform in the focal plane, producing the necessary field distribution.

Finally, we note that the estimates presented above are primarily of a qualitative nature. The exact resonance widths may vary depending on the particle geometry and illumination wavelength. The main goal of this section is to demonstrate the principal feasibility of the described effect. Moreover, in a practical implementation, some of the constraints introduced in this study can be relaxed.

**Funding.** This work was supported by a grant from the Foundation for the Development of Theoretical Physics and Mathematics «BASIS».